\documentclass[fleqn,twoside]{article}
\usepackage{espcrc2}
\usepackage{graphicx}
\usepackage[figuresright]{rotating}
\def\nabstar#1{\nabla\kern-0.5pt\smash{\raise 4.5pt\hbox{$\ast$}}
               \kern-4.5pt_{#1}}

\def\drvstar#1{\partial\kern-0.5pt\smash{\raise 4.5pt\hbox{$\ast$}}
               \kern-5.0pt_{#1}}

\def\newline{\relax\ifhmode\null\hfil\break\else\nonhmodeerr@\newline\fi}
\def\frac#1#2{{#1\over#2}}
\def\text#1{{\hbox{\rm #1}}}

\newcommand{\beq}{\begin{equation}}
\newcommand{\eeq}{\end{equation}}
\newcommand{\bea}{\begin{eqnarray}}
\newcommand{\eea}{\end{eqnarray}}
\def\Id{ \mbox{1\hspace{-1.2mm}I} }
\def\BE{\begin{equation}}
\def\EE{\end{equation}}
\def\BA{\begin{eqnarray}}
\def\EA{\end{eqnarray}}
\def\BAN{\begin{eqnarray*}}
\def\EAN{\end{eqnarray*}}

\def\gm5{\gamma_5}

\def\BE{\begin{equation}}
\def\EE{\end{equation}}
\def\BA{\begin{eqnarray}}
\def\EA{\end{eqnarray}}
\def\BAN{\begin{eqnarray*}}
\def\EAN{\end{eqnarray*}}

\def\text#1{{\rm #1}}
\newcommand{\AmS}{{\protect\the\textfont2
  A\kern-.1667em\lower.5ex\hbox{M}\kern-.125emS}}
\title{
       \vspace{-4.0cm}
       \rightline{\normalsize NTUTH-02-505F}
       \vspace{-0.2cm}
       \rightline{\normalsize August 2002}
       \vspace{2.9cm}
A computational system for lattice QCD with
overlap Dirac quarks\thanks{This work was supported in part by
National Science Council R.O.C. under the grant number NSC90-2112-M002-021}}

\author{Ting-Wai Chiu\address[PNTU]{Physics Department, National Taiwan
        University, Taipei, Taiwan 106, Taiwan.},
        Tung-Han Hsieh\addressmark[PNTU],
        Chao-Hsi Huang\addressmark[PNTU],
        Tsung-Ren Huang\addressmark[PNTU]
        }
       
\begin{document}

\begin{abstract}

We outline the essential features of a Linux PC cluster
which is now being developed at National Taiwan University,
and discuss how to optimize its hardware and software for
lattice QCD with overlap Dirac quarks. At present, the cluster
constitutes of 30 nodes, with each node consisting of one
Pentium 4 processor (1.6/2.0 GHz), one Gbyte of PC800 RDRAM,
one 40/80 Gbyte hard disk, and a network card.
The speed of this system is estimated to be 30 Gflops,
and its price/performance ratio is better than $\$1.0$/Mflops
for 64-bit (double precision) computations in
quenched lattice QCD with overlap Dirac quarks.

\vspace{1pc}
\end{abstract}

\maketitle

\section{Introduction}

It is well known that extracting physics from lattice QCD
requires computing power exceeding that of any desktop
personal computer currently available in the market.
Therefore, for one without supercomputer resources,
building a computational system seems to be inevitable
if one really wishes to pursue a meaningful
number of any physical quantity from lattice QCD.
However, the feasibility of such a project depends not only on
the funding, but also on the theoretical advancement of the subject,
namely, the realization of exact chirally symmetry on the lattice
\cite{Neuberger:1998fp,Narayanan:1995gw}.
Now, if we also take into account of the current price/performance of
PC hardware components, it seems to be the right timing to rejuvenate
the project \cite{Chiu:1988st} with a new goal - to build
a computational system for lattice QCD with exact chiral symmetry.
In this paper, we outline the essential features of a Linux PC cluster
which is now being developed at National Taiwan University,
and discuss how to optimize its hardware and software
for lattice QCD with overlap Dirac quarks.
More detailed descriptions have been given in
Ref. \cite{Chiu:2002bi}.

First, we start from quenched QCD, and compute quark propagators in the
gluon field background, for a sequence of configurations generated
stochastically with weight $ \exp( -{\cal A}_g ) $.
Then the hardronic observables such as meson and baryon correlation functions
can be constructed, and from which the hadron masses and decay constants
can be extracted. In general, one requires that any quark propagator coupling
to physical hadrons must be of the form \cite{Chiu:1998eu}
\bea
 ( D_c + m_q )^{-1} \ ,
\eea
where $ m_q $ is the bare quark mass, and
$ D_c $ is a chirally symmetric and anti-hermitian
Dirac operator\footnote{Here we assume that $ D_c $ is doubler-free,
and has correct continuum behavior, and $ D = D_c ( 1 + r a D_c )^{-1} $
is exponentially local for a range of $ r > 0 $.}.
%
For any massless lattice Dirac operator $ D $ satisfying the
Ginsparg-Wilson relation \cite{Ginsparg:1982bj}
\bea
\label{eq:gwr}
D \gamma_5 + \gamma_5 D = 2 r a D \gamma_5 D \ ,
\eea
it can be written as $ D = D_c ( 1 + r a D_c )^{-1} $ \cite{Chiu:1998gp},
and the bare quark mass is naturally added to the $ D_c $ in the numerator
\cite{Chiu:1998eu},
\BAN
D(m_q) = ( D_c + m_q ) ( 1 + r a D_c )^{-1} \ .
\EAN
Then the quenched quark propagator becomes
\BAN
( D_c + m_q )^{-1} = ( 1 - r m_q a )^{-1} [ D(m_q)^{-1} - r a ] \ .
\EAN
If we fix one of the end points at $ ( \vec{0}, 0 ) $
and use the Hermitcity $ D^{\dagger} = \gamma_5 D \gamma_5 $,
then only 12 (3 colors times 4 Dirac indices) columns of
\bea
\label{eq:propagator}
D(m_q)^{-1} =  D^{\dagger}(m_q) \{ D(m_q) D^{\dagger}(m_q) \}^{-1}
\eea
are needed for computing the time correlation functions of hadrons.
Now our problem is how to optimize a PC cluster
to compute $ D(m_q)^{-1} $ for a set of bare quark masses.

For overlap Dirac quarks, we need to solve the following linear system
by (multi-mass) conjugate gradient (CG),
\bea
\label{eq:outer_CG}
D D^{\dagger} Y
= \left( \alpha + \eta P_{\pm} H_w \frac{1}{\sqrt{H_w^2}} \right) Y = \Id
\eea
where $ P_{\pm} = ( \gamma_5 \pm 1 )/2 $, $ \alpha = 2 m_0^2 + m_q^2/2 $,
$ \eta = 2 m_0^2 - m_q^2/2 $, and $ m_0 $ is fixed to $ 1.3 $ in our
computations.
Then the quark propagators can be obtained via (\ref{eq:propagator}).
With Zolotarev optimal rational approximation
\cite{Zolotarev:1877,Akhiezer:1992}
to $ (H_w^2)^{-1/2} $, the multiplication
($ h_w \equiv H_w / \lambda_{min} $)
\bea
H_w \frac{1}{\sqrt{H_w^2}} Y
 \simeq h_w ( h_w^2 + c_{2n} )
 \sum_{l=1}^{n} \frac{ b_l }{ h_w^2 + c_{2l-1} } Y
\label{eq:mult_Y}
\eea
can be evaluated by invoking another (multi-mass)
conjugate gradient process to the linear systems
\bea
\label{eq:inner_CG}
( h_w^2 + c_{2l-1} ) Z_l = Y, \hspace{4mm} l=1, \cdots, n \ .
\eea
where the coefficients $ c_l $, $ b_l $ and $ d_0 $
have been given explicitly in Ref. \cite{Chiu:2002eh}.

In order to improve the accuracy of the rational approximation
as well as to reduce the number of iterations in the inner CG loop,
it is {\it crucial} to project out the largest and some low-lying
eigenmodes of $ H_w^2 $. Denoting these eigenmodes by
\bea
\label{eq:eigen}
H_w u_j = \lambda_j u_j, \hspace{4mm} j = 1, \cdots, k ,
\eea
then we project the linear systems (\ref{eq:inner_CG}) to the
complement of the vector space spanned by these eigenmodes
\bea
\label{eq:inner_CG1}
( h_w^2 + c_{2l-1} ) \bar{Z_l} = \bar{Y}
\equiv ( 1 - \sum_{j=1}^k u_j u_j^{\dagger} ) Y .
\eea
Thus the r.h.s. of (\ref{eq:mult_Y}) can be rewritten as
\bea
\label{eq:SS}
h_w ( h_w^2 + c_{2n} ) \sum_{l=1}^{n} b_l \bar{Z}_l +
\sum_{j=1}^k \frac{\lambda_j}{\sqrt{\lambda_j^2}} u_j u_j^{\dagger} Y \equiv S
\eea
Then the breaking of exact chiral symmetry (\ref{eq:gwr}) can be measured
in terms of
\bea
\label{eq:sigma}
\sigma = \frac{ | S^{\dagger} S - Y^{\dagger} Y | }{  Y^{\dagger} Y } \ .
\eea
In practice, one has no difficulties to attain $ \sigma < 10^{-12} $
for most gauge configurations on a finite lattice (Table \ref{tab:perf2}).

Now the computation of overlap Dirac quark propagator involves
two nested conjugate gradient loops: the so-called inner CG loop
(\ref{eq:inner_CG1}), and the outer CG loop (\ref{eq:outer_CG}).

\begin{table*}[htb]
\caption{
  The execution time (in unit of second) of a Pentium 4 (2 GHz) node to
  compute 12 columns of overlap Dirac quark propagators,
  versus the size of the lattice.
  The parameters for the test are:
  the degree of Zolotarev rational polynomial is $ n=16 $,
  the number of bare quark masses is $ N_m=16 $ (with $ ma \ge 0.02 $),
  the precision of each projected eigenmode satisfies
             $\|(H_w^2-\lambda^2) |x\rangle \|< 10^{-13}$, and
  the stopping criterion for the inner and outer CG loops is
  $ \epsilon=10^{-11} $.}
\begin{center}
\begin{tabular}{c|cccccccccc}
\hline
& & \multicolumn{1}{c}{project.} & & $\chi$ sym& inner CG & outer CG &
        disk I/O & Total \\
Lattice & $\beta$ & no./time & $ \lambda_{min}$/$\lambda_{max}$ &
        $\sigma$(max.) & ave. iters. & tot. iters.
        & time & time \\
\hline\hline
$ 8^3\times 24$ & 5.8 & 32/4725 & 0.198/6.207
        &  $ 5.3 \times 10^{-14} $ &
        403 & 1282 & 0  & 63550 \\
$10^3\times 24$ & 5.8 & 30/7803 & 0.152/6.204 &
        $ 6.4 \times 10^{-14}$ &
        519 & 1943 & 0 & 218290 \\
$12^3\times 24$ & 5.8 & 30/13258 & 0.129/6.211 &
        $ 9.8 \times 10^{-14}$ &
        608 & 2840 & 0 & 625718 \\
$16^3\times 32$ & 6.0 & 20/74937 & 0.215/6.260 & $ 3.3 \times 10^{-13}$
        &  370 & 3968 & 66976 & 2095975 \\
\hline
\end{tabular}
\label{tab:perf2}
\end{center}
\end{table*}

\section{Optimization}

Next we address the question how to configure the hardware and software
of a PC cluster such that it can attain the optimal price/performance
ratio for the execution of the nested CG loops.

The vital observation is that {\it not} all column vectors are used
simultaneously at any step of the nested CG loops, and also the
computationally intense part is at the inner CG loop (\ref{eq:inner_CG1}).
Thus we can use the hard disk as virtual memory for the storage of
the intermediate solution vectors and their conjugate gradient vectors
at each iteration of the outer CG loop, while the CPU is
working on the inner CG loop. Then the amount of required memory at
each node can be greatly reduced. It is easy to figure out \cite{Chiu:2002bi}
the minimum memory required for accommodating the link variables and
all relevant vectors for the inner CG loop,
\bea
\label{eq:Ncg_min}
N_{cg}^{min} = 384 \times N_s ( n + 3 ) \ \mbox{byte} \ ,
\eea
where $ n $ is the degree of the Zolotarev rational polynomial, and
$ N_s $ is the total number of lattice sites.
For computations of quark propagators
on the $ 16^3 \times 32 $ lattice
with $ n=16 $, (\ref{eq:Ncg_min}) gives
$ N_{cg}^{min} = 0.912 \ \mbox{Gbyte} $,
which can be implemented in a single node with one Gbyte
of memory (i.e., four stripes of 256 Mbyte memory modules).
Then the maximum speed of the cluster can be attained since
there is no communication overheads.
Moreover, the time for disk I/O
(at the interface of inner and outer CG loops)
only constitutes a few percent of the total time for the
entire nested CG loops (Table \ref{tab:perf2}).
Further, to take advantage of the vector unit of Pentium 4, we
rewrite the computationally intense part
($ H_w $ times a vector $ Y $)
of our program in SSE2 codes\cite{r:intel:sse2,Luscher:2001tx,Chiu:2002bi},
which yields a speed-up by a factor of $\simeq 1.8$.

\section{Conclusions}

The speed of our system of 30 nodes is higher than 30 Gflops,
and the total cost of the hardware is less than US\$30,000.
This amounts to price/performance ratio better than \$1.0/Mflops for
64-bit (double precision) computations with overlap Dirac quarks.
The basic idea of optimization is to let each node compute one of
the 12 columns of the quark propagators (for a set of bare quark masses),
and also use the hard disk as virtual memory for the vectors
in the outer CG loop, while the CPU is working on the inner CG loop.

With Zolotarev optimal rational approximation to $ ( H_w^2 )^{-1/2} $,
projection of high and low-lying eigenmodes of $H_w^2$,
the multi-mass CG algorithm, the SSE2 acceleration,
and the memory management,
we are able to compute overlap Dirac quark propagators
of $ 16 $ bare quark masses on the $16^3 \times 32$ lattice,
with the precision of quark propagators up to $ 10^{-11} $
and the precision of exact chiral symmetry up to $ 10^{-12} $,
at the rate of two gauge configurations ( $ \beta = 6.0 $ ) per two days,
with our present system of 30 nodes (Table \ref{tab:perf2}).
This demonstrates that an optimized Linux PC cluster
can be a viable computational system to extract
physical quantities from lattice QCD
with overlap Dirac quarks \cite{Chiu:2002xm}.

\end{document}